\documentclass[aps,prc,twocolumn,superscriptaddress]{revtex4-1}

\usepackage{graphicx}
\usepackage{longtable}
\usepackage{rotating}

\begin{document}


\title{Structure of $^8$B from elastic and inelastic $^7$Be+p scattering}



\author{J.P. Mitchell}
\email{jmitchel@astro.puc.cl}
\affiliation{Department of Physics, Florida State University, 32306 FL}
\affiliation{Departmento de Astronom\'{i}a y Astrof\'{i}sica, Pontificia Universidad Cat\'{o}lica de Chile, Vicu\~{n}a Mackenna 4860, Macul, Santiago, Chile}
\affiliation{Argelander Institut f\"{u}r Astronomie, Universit\"{a}t Bonn, Auf dem H\"{u}gel 71, 53121 Bonn, Germany}

\author{G.V. Rogachev}
\email{grogache@fsu.edu}
\affiliation{Department of Physics, Florida State University, 32306 FL}
\affiliation{National Superconducting Cyclotron Laboratory, Michigan State University, East Lansing, MI 48824}

\author{E.D. Johnson}
\affiliation{Department of Physics, Florida State University, 32306 FL}
\author{L.T. Baby}
\affiliation{Department of Physics, Florida State University, 32306 FL}
\author{K.W. Kemper}
\affiliation{Department of Physics, Florida State University, 32306 FL}
\author{A.M. Moro}

\affiliation{Department of Physics, University of Seville, Spain} 

\author{P. Peplowski}
\affiliation{Department of Physics, Florida State University, 32306 FL}
\affiliation{Johns Hopkins University Applied Physics Laboratory, Laurel, MD 20723}

\author{A. Volya}
\affiliation{Department of Physics, Florida State University, 32306 FL}

\author{I. Wiedenh\"over}
\affiliation{Department of Physics, Florida State University, 32306 FL}



\date{\today}

\begin{abstract}
\begin{description}
\item[Motivation] Detailed experimental knowledge of the level structure of light weakly bound nuclei is necessary to guide the development of new theoretical approaches that combine nuclear structure with reaction dynamics.
\item[Purpose] The resonant structure of $^8$B is studied in this work.
\item[Method] Excitation functions for elastic and inelastic $^7$Be+p scattering were measured using a $^7$Be rare isotope beam. Excitation energies ranging between 1.6 and 3.4 MeV were investigated. An R-matrix analysis of the excitation functions was performed. 
\item[Results] New low-lying resonances at 1.9, 2.5, and 3.3 MeV in $^8$B are reported with spin-parity assignment 0$^+$, 2$^+$, and 1$^{+}$, respectively. Comparison to the Time Dependent Continuum Shell (TDCSM) model and {\it ab initio} no-core shell model/resonating-group method (NCSM/RGM) calculations is performed. This work is a more detailed analysis of the data first published as a Rapid Communication.  [J.P. Mitchell, et al, Phys. Rev. C 82, 011601(R) (2010)]
\item[Conclusions] Identification of the 0$^+$, 2$^+$, 1$^{+}$ states that were predicted by some models at relatively low energy but never observed experimentally is an important step toward understanding the structure of $^8$B. Their identification was aided by having both elastic and inelastic scattering data.  Direct comparison of the cross sections and phase shifts predicted by the TDCSM and {\it ab initio} No Core Shell Model coupled with the resonating group method is of particular interest and provides a good test for these theoretical approaches.
\end{description}
\end{abstract}


\maketitle

\section{Introduction}

\begin{figure}
\includegraphics[width=1.0\linewidth]{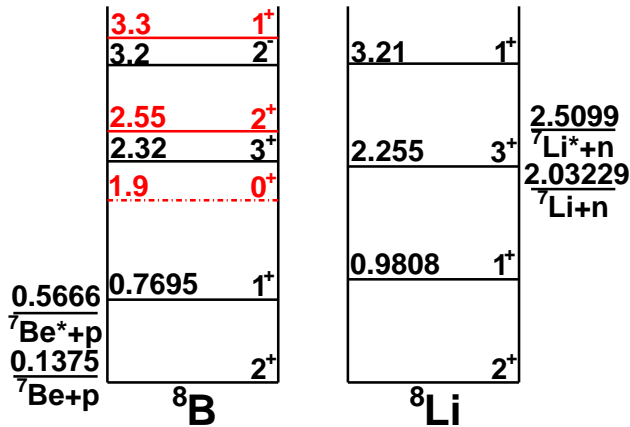}
\caption{\label{fig:levelscheme} (Color online)  The level schemes for $^{8}$B and its mirror $^{8}$Li.  States from our previous work \cite{Mitchell10} are in red.  The dashed-dotted line indicates that the state is tentative.} 
\end{figure}

One of the main goals of modern nuclear theory is to combine the nuclear reaction models with nuclear structure calculations to provide the unified framework that allows the calculation of level spectroscopy and reaction cross sections starting from the same established nuclear Hamiltonian. Several theoretical approaches have been suggested to advance this goal. Broadly, two major directions can be identified, phenomenological and {\it ab-initio}. The first one uses the well established shell-model Hamiltonian and couples it with the corresponding reaction channels. The recoil corrected continuum shell model (RCCSM) \cite{Halderson04} and the time-dependent continuum shell model (TDCSM) \cite{Volya2009} are examples of these approaches. The second major direction is the attempt to calculate the cross section starting from both bare nucleon-nucleon forces and three-nucleon forces. One example of this approach is the no-core shell model combined with the resonating-group method (NCSM/RGM) \cite{Navratil11}. The very attractive feature of these developments is that the excitation functions of the resonance reactions, such as elastic and inelastic nucleon scattering, (p,n) and (p,$\alpha$) reactions, etc., can, in principle, be calculated and directly compared to the experimental data. This is in addition to all known structure data.  However, this comparison is not as straightforward as it may appear. Because of model space truncation, limitations from nucleonic degrees of freedom and numerical complexity, it is natural to expect that the nuclear spectrum at the low excitation energy is reproduced better than the spectrum of the higher lying excited states by any model.  Therefore, it is desirable to verify the theoretical predictions in the region of low excitation energy first and 
weakly bound nuclei provide a good test for these models.  Here, the continuum appears at low energy, thus permitting examination of the structure-reaction transition.  Moreover, because of truncation of the model space, parameters of models are adjusted to the well known spectrum of stable nuclei, resulting in unsurprisingly reasonable agreement with the experimental data for these nuclei.  The better test is provided by exotic, weakly bound nuclei. The neutron deficient Boron isotope, $^8$B, is a prime example. Its proton separation energy is only 137 keV and all of its excited states are in the continuum, as can be seen in its level structure in Fig. \ref{fig:levelscheme}. In addition, this nucleus has been a subject of numerous theoretical studies.  In the recent {\it ab initio} NCSM/RGM analysis \cite{Navratil11} of $^8$B the proton+$^7$Be elastic scattering phase shifts as well as the  cross section for the $^7$Be(p,p') and the $^7$Be(p,$\gamma$) reactions were calculated. Direct comparison of the experimental results on the $^7$Be(p,p) and $^7$Be(p,p') reactions with these calculations and also the analysis of the experimental data using the TDCSM approach is the main objective of this work. 

The excitation function for $^7$Be+p has been previously measured in several experiments \cite{Goldberg98,Rogachev01,Angulo03,Yamaguchi09}. However, the thick target inverse kinematics experimental method used in all of these measurements did not allow for separation between elastic and inelastic scattering except for the data from \cite{Angulo03}, where measurements were performed at energies below the inelastic scattering threshold. In Ref. \cite{Yamaguchi09}, an attempt has been made to use $\gamma$-proton coincidence to identify the inelastic scattering events, however, the elastic excitation function still appears to be contaminated with inelastic events (see section \ref{sec:Experimental-Setup} for additional comments). The experiment described here does not suffer from such deficiency because the Intermediate Target Thickness Approach has been applied. This approach allowed for measurement of a significant fraction of the $^{7}$Be+p excitation function, while simultaneously detecting the $^{7}$Be recoil in coincidence with protons in order to distinguish between elastic and inelastic scattering events kinematically. Therefore, we did not use experimental data from the previous higher energy measurements \cite{Goldberg98,Rogachev01,Yamaguchi09} in the analysis but we included the low energy $^7$Be(p,p) elastic scattering data between 0.3 and 0.75 MeV measured in Ref. \cite{Angulo03}. The subset of the data reported here was first published in \cite{Mitchell10}.  

This paper contains a more detailed description of the experimental results and also extends the previously measured excitation energy region to higher energies. A description of the experimental method that was used to measure the excitation functions for $^7$Be+p elastic and inelastic scattering between 1.6 to 3.4 MeV is given in section \ref{sec:Experimental-Setup}. The analysis of the experimental data was performed using the multi-channel multi-level R-matrix approach and is described in section III. 
Section IV contains a discussion of this finding and its consistency with the previous experimental data on the $^8$B and $^8$Li nuclei and discusses whether it is possible to explain the new experimental data without introducing the new resonances in $^8$B. Analysis of the new experimental data in the framework of the Time Dependent Continuum Shell Model is presented in section V. Detailed comparison of the phase shifts extracted from the analysis of the p+$^7$Be experimental data to the predictions of the {\it ab initio} NCSM/RGM model is given in section VI. Conclusions are given in Chapter VII.

\begin{figure}
\includegraphics[width=0.85\linewidth]{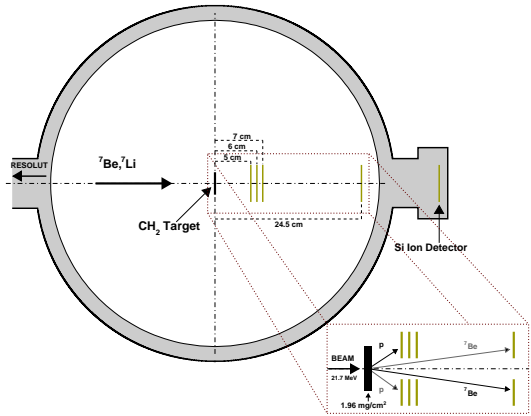}
\caption{\label{fig:setup}(Color online) The experimental setup. The $^{7}$Be beam was delivered by the RESOLUT facility (on the left).  The C$_{2}$H$_{4}$ targets of various thicknesses were used.  The protons were detected in an array of three Micron Semiconductors S2 detectors and the $^{7}$Be's were measured in an S2 downstream.  (The inset provides a more detailed view of the detector arrangement.)}
\end{figure}

\section{\label{sec:Experimental-Setup}Experiment}

The excitation function for p+$^7$Be elastic and inelastic scattering between 1.6 and 3.4 MeV in c.m.s. was measured at the John D. Fox Superconducting Accelerator Laboratory at Florida State University. A radioactive beam of $^{7}$Be was produced using the $^{1}$H($^{7}$Li,$^{7}$Be)n reaction. A primary $^{7}$Li beam was accelerated by a 9MV SuperFN Tandem Van de Graaff accelerator followed by a LINAC booster. The primary target was a 4 cm long hydrogen gas cell with 2.5 $\mu$m Havar entrance and exit windows. The gas cell was cooled by liquid nitrogen and had a gas pressure of 390 mBar. The in-flight production rare isotope beam facility RESOLUT was used to separate $^{7}$Be from other reaction products and the primary beam. RESOLUT is a set of two superconducting solenoids, dipole and quadrupole magnets and a superconducting resonator. Three $^{7}$Be beam energies were used in this experiment: 27.2, 22.0, and 18.5 MeV. The typical intensity of the $^{7}$Be beam was $10^{5}$ pps. The composition of the beam was 70$\%$ $^{7}$Be and 30$\%$ $^{7}$Li contaminant. Diagnostics of the secondary beam were performed using a position sensitive Micro-Channel Plate detector installed between the dipole magnet and the second solenoid (2.7 m before the C$_{2}$H$_{4}$ target) and the $\Delta$E-E telescope consisting of an ionization chamber (used as $\Delta$E detector) backed by a 50x50 mm$^{2}$ 16x16 Silicon strip detector positioned 66 cm downstream from the secondary target.

\begin{figure}
\includegraphics[width=1.0\linewidth]{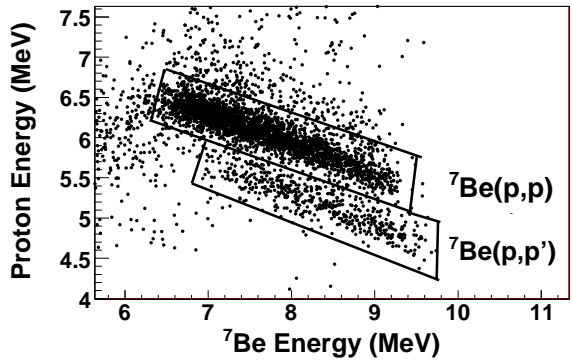}
\caption{\label{fig:2d}Scatter plot of kinematic coincidence between protons and $^7$Be ions. Regions which correspond to elastic and inelastic scattering are labeled.} 
\end{figure}

A sketch of the experimental setup is shown in Fig. \ref{fig:setup}. A solid polyethylene (C$_{2}$H$_{4}$) target of thickness optimized for the given beam energy (see description below) was used. A set of three annular Micron Semiconductor silicon strip detectors (S2 design) for the proton recoils were positioned 5, 6 and 7 cm downstream from the target, respectively. Another S2 detector for the $^{7}$Be recoils was positioned 24.5 cm from the target. The S2 detector has annular geometry and consists of 16 segments and a side of rings that allow for the scattering angle of the products to be determined. The first in the set of three proton detectors was a $\Delta$E detector of 65 $\mu$m, while the other two and the $^{7}$Be detector were 500 $\mu$m each.

The target thickness was optimized for maximum energy losses of the $^{7}$Be ions in the target while ensuring that all $^{7}$Be recoils make it out of the target with enough kinetic energy left to be detected in the downstream S2 detector. Kinematic coincidence between protons in the array of three S2 detectors and the $^{7}$Be recoils in the downstream S2 detector were then used to identify the scattering events. The 65 $\mu$m $\Delta$E S2 detector was used only in the initial stage of the experiment to verify that kinematic coincidence between light and heavy recoils are enough for clean identification of the p+$^{7}$Be elastic and inelastic scattering events. This detector was then removed.  Measurements at the beam energies of 22 and 18.5 MeV were performed without the 65 $\mu$m detector, while that for the 27.2 MeV energy included the 65 $\mu$m $\Delta$E detector. Time between the events in the proton and $^{7}$Be detectors was measured with resolution of about 3 ns in order to eliminate random coincidence background. Elastic and inelastic scattering processes can be distinguished, because complete kinematics of the events are measured. More specifically, events that have two equal energy protons would have different energy of $^{7}$Be recoils if they originate from different (elastic/inelastic) processes. This is due to different reaction Q-value and kinematics, and also effective target thicknesses (and hence energy losses) experienced by the heavy recoils. The inelastic events that produce protons with the same kinetic energy as elastic events take place earlier (upstream) in the target, where a negative reaction Q-value is compensated by the higher energy of the $^{7}$Be projectile. (See also Ref. \cite{Rogachev2010} for details on this experimental technique). The 2D scatter plot for the kinematic coincidence between protons and $^{7}$Be is shown in Fig. \ref{fig:2d}. The kinematic loci which correspond to elastic and inelastic scattering processes are labeled and outlined with contours. Polyethylene target thicknesses used in this experiment were 2.6, 2.5 and 1.5 mg/cm$^{2}$ for the 27.2, 22 and 18.5 MeV beam energies respectively. In addition, a separate run at 18.5 MeV of $^{7}$Be beam energy was performed with a slightly thicker (2 mg/cm$^{2}$) target, to extend the measured excitation function to lower energies without changing the energy of the beam. Under this condition coincidence between the highest energy protons and the $^{7}$Be recoils are lost (the heavy recoils produced at the beginning of the target do not make it through). Only the lower energy part of this spectrum was used in the analysis.

\begin{figure}
\includegraphics[width=1.0\linewidth]{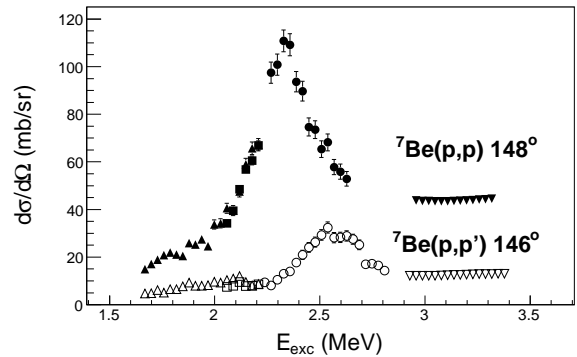}
\caption{\label{fig:combined}The excitation function for $^7$Be+p elastic and inelastic scattering at 148$\pm$4$^{\circ}$ and 146$\pm$4$^{\circ}$ degrees respectively. Results from runs at three different energies of $^7$Be beam are shown. The squares correspond to the run at 18.5 MeV of $^7$Be with a 1.5 mg/cm$^{2}$ target, the vertex up triangles are data taken at 18.5 MeV with the 2 mg/cm$^{2}$ target, the circles are the 22 MeV data with the 2.5 mg/cm$^{2}$ target, and the vertex down triangles are from the 27.2 MeV run with a 2.6 mg/cm$^{2}$ with solid markers representing $^{7}$Be+p elastic scattering and open markers the inelastic p($^{7}$Be,p')$^{7}$Be($\frac{1}{2}^{-}$) scattering excitation functions.} 
\end{figure}

\begin{figure}
\includegraphics[width=1.0\linewidth]{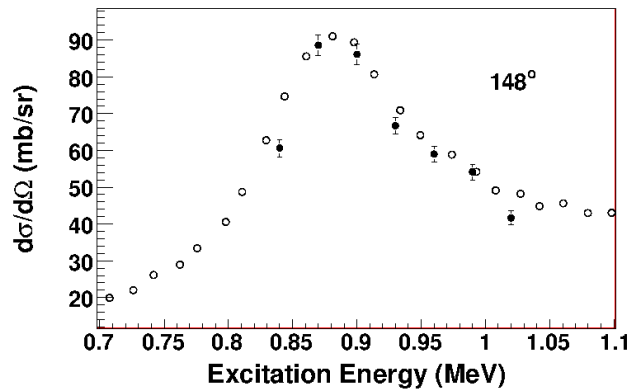}
\caption{\label{fig:7Lipnorm}The excitation function for $^7$Li+p elastic scattering at 148$\pm$4$^{\circ}$ is shown with solid circles.  This excitation function was measured simultaneously with $^7$Be+p (the rare isotope beam composition was 70$\%$ $^{7}$Be and 30$\%$ $^{7}$Li) and used for absolute normalization.  The same excitation function from \cite{Bashkin1951,Walters1953} is shown for comparison with open circles.} 
\end{figure}

Fig. \ref{fig:combined} shows excitation functions for resonance elastic and inelastic scattering of $^{7}$Be+p measured in four different runs. The vertex up triangles correspond to the $^{7}$Be run at 18.5 MeV with the 2 mg/cm$^{2}$ target, the squares are 18.5 MeV $^{7}$Be with the 1.5 mg/cm$^{2}$ target data, the circles are 22 MeV $^{7}$Be with 2.5 mg/cm$^{2}$ data, the vertex down triangles are from the 27.2 MeV run with the 2.6 mg/cm$^{3}$ target, and in all cases, the filled markers are for elastic scattering and the hollow markers for inelastic scattering. The angular resolution of the experimental setup, as determined by the pitch of the rings in the S2 detector, distance from the target and the size of the beam spot on the secondary target, was 1.25$^{\circ}$. We used binning of 4$^{\circ}$ in the lab frame, combining events recorded by 12 rings of the S2 detector into one spectrum. Excitation functions at three scattering angles were obtained this way. These angles are 148$\pm$4$^{\circ}$, 140$\pm$4$^{\circ}$ 132$\pm$4$^{\circ}$ in c.m.s. for elastic scattering and 146$\pm$4$^{\circ}$, 138$\pm$4$^{\circ}$ 130$\pm$4$^{\circ}$ for inelastic scattering. Absolute normalization of the cross section was performed using the known excitation functions for $^{7}$Li+p elastic scattering. These excitation functions were extracted from the experimental data using the same procedure as for the $^{7}$Be+p elastic scattering, therefore, by normalizing the $^{7}$Li+p data to the known $^{7}$Li+p cross section and taking into account the ratio of the $^{7}$Be ions to the $^{7}$Li ions in the secondary beam (as measured by the 0 degree ionization chamber and silicon strip detector), accurate normalization is achieved. Note that this normalization procedure automatically takes into account the efficiency of the experimental setup. A sample of the $^{7}$Li+p excitation function measured in this experiment is shown in Fig. \ref{fig:7Lipnorm} (solid circles) and compared to the experimental data from \cite{Bashkin1951,Walters1953}. Excitation functions extracted from our data agree well with the differential cross section for elastic and inelastic scattering of $^{7}$Be+p measured at several energies of $^{7}$Be using a thin target approach and reported by U. Greife, et al., \cite{Greife2007}.  The excitation functions of Yamaguchi et al. \cite{Yamaguchi09} however, differ from ours, especially in the inelastic channel where they found the excitation function to be fairly flat across their entire energy range measured, while our results have a large peak at an excitation energy of 2.5 MeV. This discrepancy may be related to the background in the NaI scintillator detectors used in Ref. \cite{Yamaguchi09} that could have prevented a clean $\gamma$-proton coincidence spectrum to be extracted.

\section{\label{sec:Rmatrix} R-matrix analysis}
\begin{figure}
\includegraphics[width=1\columnwidth]{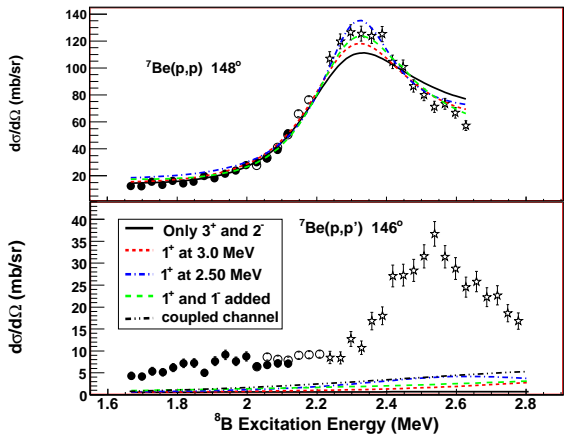} 
\caption{\label{fig:nonew}(Color Online)  R-matrix fit of the elastic and inelastic $^7$Be+p scattering with known 3$^{+}$ and 2$^{-}$ states, a second excited 1$^+$ seen in $^{8}$Li, and the ``background'' 1$^-$ state introduced at higher energy. The solid curve corresponds to only 3$^+$ and 2$^-$ states at 2.3 and 3.5 MeV, respectively. The red short-dashed curve includes the contribution of the higher lying 1$^+$ states assumed at 3.0 MeV. Dash-dotted purple curve shows the 1$^+$ state shifted to 2.5 MeV and the long dashed green curve also includes the 1$^-$ state introduced at 5 MeV.}
\end{figure}

The excitation functions for elastic $^{1}$H($^{7}$Be,p)$^{7}$Be(g.s.) and inelastic $^{1}$H($^{7}$Be,p')$^{7}$Be($1/2^{-}$; 0.43 MeV) scattering were analyzed using a two channel, multi-level R-matrix approach. The natural starting point for the analysis is to introduce only the excited states of $^8$B that were identified in previous experiments \cite{Tilley04}, the 1$^+$ at 0.77 MeV, the 3$^+$ at 2.32 MeV and the broad 2$^-$ at $\sim$3 MeV. These three states reproduce the excitation function for p+$^7$Be elastic scattering between 0.5 and 3.5 MeV reasonably well, as shown in Fig. \ref{fig:nonew} (top) by the solid line. However, it is not possible to explain 30 mb/sr inelastic cross section at 2.5 MeV if only known states in $^{8}$B are considered (Fig. \ref{fig:nonew} (bottom)). This failure can be understood from the following simple considerations. The first excited 1$^{+}$ state at 0.77 MeV is too narrow to have any significant impact on the excitation functions at energies above 1.6 MeV. The second excited state, 3$^{+}$ at 2.32 MeV, can only decay to the $3/2^-$ ground state of $^7$Be because decay to the $1/2^{-}$ first excited state requires angular momentum of $\ell=3$. Therefore, even if the corresponding reduced width is large the inelastic partial proton width, $\Gamma_{p'}=2P_{\ell}(kR)\gamma^{2}$, would still be small compared to the elastic partial proton width due to a small penetrability factor for high angular momentum decay. Hence, the cross section for population of the first excited state in $^{7}$Be due to the 3$^{+}$ resonance in $^{8}$B, determined by the $\Gamma_{p}\Gamma_{p'}/\Gamma_{\rm tot}^{2}$ ratio, is small. The same is true for the broad 2$^{-}$ state in $^{8}$B at $\approx$3 MeV as it can only decay to the first excited state in $^{7}$Be with angular momentum $\ell=2$ while decay to the g.s. proceeds with $\ell=0$. Fig. \ref{fig:nonew} shows the results of an R-matrix calculation with only previously known 1$^{+}$, 3$^{+}$ and 2$^{-}$ states at 0.77, 2.32 and 3.7 MeV with reduced width parameters evaluated using the TDCSM (more details on TDCSM calculations are given in Section \ref{sec:csm}) and known total widths of these states. (Excitation energy and width of the 2$^-$ were adjusted slightly to produce a better fit.). It is clear that while the elastic scattering data is well reproduced, the inelastic scattering data cannot be explained by the known states.

Based on the level scheme of $^{8}$Li (Fig. \ref{fig:levelscheme}) it is natural to introduce the second 1$^{+}$ state in $^{8}$B at an excitation energy around 3 MeV. Reduced widths for this state were chosen according to TDCSM calculations carried out with the Cohen-Kurath CKI interaction \cite{Cohen1965}. It was verified that these reduced widths reproduce the known width of this state in $^8$Li ($\sim$1 MeV). The short dashed curve (red) in Fig. \ref{fig:nonew} shows the effect of the 1$^+$ state on the fit. While the elastic excitation function is fitted well, the inelastic cross section is still underestimated. Even if this state is shifted to 2.5 MeV, where inelastic scattering has its maximum cross section, it still underestimates the data (dash-dotted (blue) curve in Fig. \ref{fig:nonew}). Finally, in an attempt to increase the inelastic cross section without using new states below 3 MeV we introduced a $1^{-}$ ``background'' state. This state can decay to the first excited state of $^7$Be with $\ell=0$, therefore it may contribute significantly to the inelastic cross section. The reduced widths for the 1$^-$ state were evaluated using the shell model, and the state was introduced at 5 MeV. As expected, the 1$^-$ background state increased the inelastic cross section overall (long-dashed green curve in Fig. \ref{fig:nonew}). But even with this state included the inelastic cross section cannot be reproduced.

\begin{figure*}
 \includegraphics[width=1.0\linewidth]{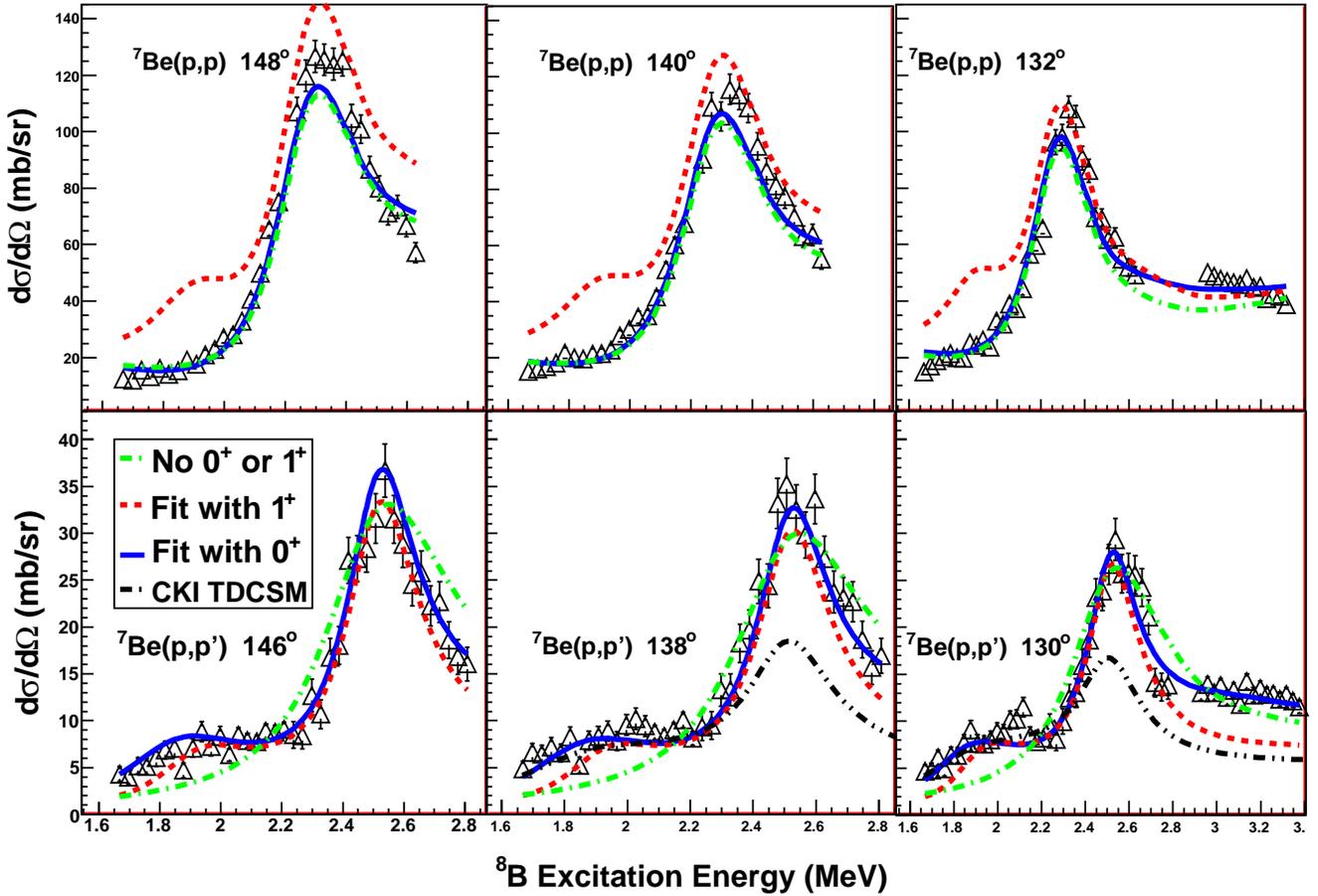}
\caption{\label{fig:final}(Color online)  Elastic and inelastic excitation functions for $^{7}$Be+p scattering.  The green dashed-dotted curve is a fit with the previously known 1$^{+}$, 3$^{+}$, and 2$^{-}$ states, as well as a 2$^{+}$ at 2.54 MeV to reproduce the peak in the inelastic data.  The red dashed curve is a fit with the aforementioned states and the 1$^+_2$ state. The blue solid curve is the best fit with all states from Table \ref{tab:states}.  The black dashed-dotted curve is a TDCSM calculation.
}
\end{figure*}

The {\it ab initio} calculations for $^8$B \cite{Navratil98,Wiringa00,Navratil10,Navratil11} predict three more positive parity (p-shell) states at low excitation energy. These are the  0$^+$$_1$, 1$^+$$_2$ and 2$^+$$_2$. The excitation energies for these states vary between 2 and 6 MeV depending on the three-body force parameterization and the specifics of the calculations. Similar results are obtained in shell model calculations (excitation energies of these ``missing'' states vary between 2 and 6 MeV in the shell model as well, depending on the residual interaction used). Therefore, it is natural to introduce these states in an attempt to reproduce the large inelastic scattering cross section. The 1$^+$$_2$ state has already been introduced. That leaves only the 0$^+$$_1$ and 2$^+$$_2$. Introduction of a new 2$^{+}$ state placed at 2.5 MeV, reproduces both the magnitude and angular dependence of the observed peak in the inelastic cross section while keeping the elastic excitation function in agreement with the experimental data (green dash-dotted curve in Fig. \ref{fig:final}). However, even with this new state the cross section for inelastic scattering below 2.3 MeV is still underestimated. The 2$^{+}$ state should have a relatively small width (270$\pm$40 keV) to fit the observed peak-like structure in the inelastic excitation function at 2.5 MeV and its influence below 2.3 MeV is small. Introducing the 0$^{+}$ state at an excitation energy of 1.9$\pm$0.1 MeV with a width of 530$_{-100}^{+600}$ keV allows the inelastic scattering data to be fit below 2.3 MeV without destroying the fit to the elastic scattering data (solid blue line in Fig. \ref{fig:final}). It was verified that a 1$^{+}$ spin-parity assignment for this state is not possible as it will ruin agreement with the elastic scattering data (red dashed line in Fig. \ref{fig:final}).  

\begin{figure}
 \includegraphics[width=1.0\linewidth]{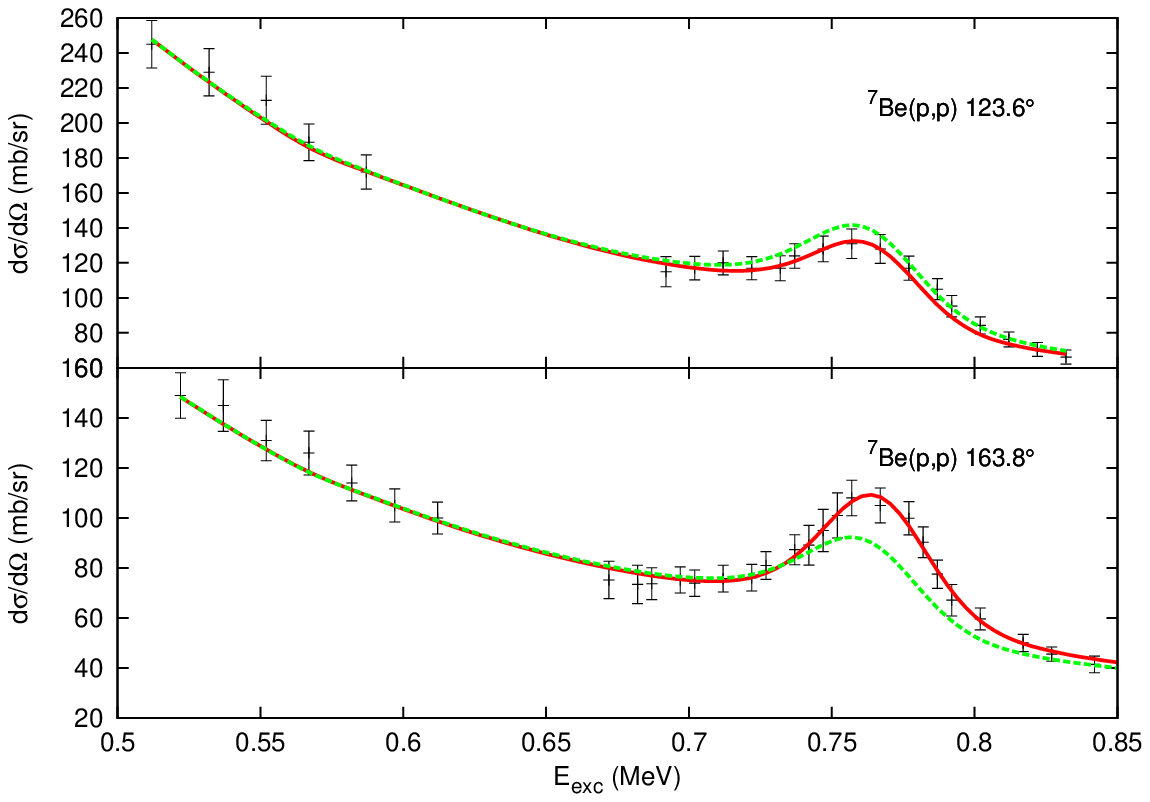}
\caption{\label{fig:angulofig} (Color online) Excitation function for $^7$Be+p elastic scattering at low energies from \cite{Angulo03} at 123.6$^{\circ}$ and 163.8$^{\circ}$. The best fit is a solid red curve. The calculated cross section was convoluted to account for 30 keV experimental resolution reported in Ref. \cite{Angulo03}. Systematic errors were included into the error bars. The dashed green curve is the R-matrix fit with the {\it ab initio} 1$^+$ phase shifts from \cite{Navratil10} for the 1$^+_1$ state (channel spins 1 and 2 contribute about equally).}
\end{figure}

The low-energy data from \cite{Angulo03} were used (Fig. \ref{fig:angulofig}) to provide additional constrain on the behavior of the phase shifts at low energy. It proved to be particularly important for the negative parity phase shifts. We used the predictions of the {\it ab initio} calculations \cite{Navratil10} for the 2$^-$ and 1$^-$ phase shifts as the starting point, but the best fit was achieved with the negative parity phase shifts different from \cite{Navratil10}. (It is discussed in more detail in Chapter \ref{abinitio}.) The best fit that included the low energy data from \cite{Angulo03} and data from this experiment was achieved using R-matrix parameters given in Table \ref{tab:states}. The normalized $\chi^2$ for the best fit was 0.89. States shown in parenthesis in Table \ref{tab:states} are the broad ``background'' states that are used in R-matrix formalism to produce the desired behavior of the corresponding phase shifts.

\begin{table*}
\caption{\label{tab:states} Parameters of resonances in $^8$B from the R-matrix best fit. States in parenthesis are outside of the measured excitation energy range but provide essential ``background'' through low energy tails.  The energy eigen-value and the  reduced widths amplitudes for $^7$Be(p,p) and $^7$Be(p,p')$^7$Be(1/2$^-$) scattering with channel spins 1 and 2 for the former and 0 and 1 for the latter used in the R-matrix fit are shown in columns 6-10.  We used 4.20 fm as the channel radius for both the elastic S=1,2 and inelastic S=0,1 channels.}
\begin{ruledtabular}
\begin{tabular}{ccccccccccc}
J$^{\pi}$ & E$_{ex}$ (MeV)& $\Gamma_{tot}$ (MeV)& $\Gamma_{p}$ (MeV)& $\Gamma_{p'}$ (MeV)& E$_{eigen}$ & $\gamma_{el}$ S=1 & $\gamma_{el}$ S=2 & 
$\gamma_{1/2^-}$ S=0 & $\gamma_{1/2^-}$ S=1 \\
\hline
$2^{+}$ &    0       &    -     &    -     &   -   & -0.657 & -0.793 & -0.531 & 0.000 & 0.430 \\
$1^{+}$ & 0.768(4) & 0.027(6) & 0.026(6) & 0.001 & 0.276 &  0.718 & 0.130 & -0.875 & -0.335 \\
$0^{+}$ & 1.9(1) & 0.53$_{-0.1}^{+0.6}$ & 0.06$_{-0.02}^{+0.3}$ & 0.47$_{-0.1}^{+0.4}$ &  2.102 & 0.353 & 0.000 & 0.000 & 1.303 \\
$3^{+}$ & 2.31(2) & 0.33(3) & 0.33(3) & 0.0 &  2.305 & 0.000 &  0.607 & 0.000 & 0.000 \\
$2^{+}$ & 2.50(4) & 0.27(4) & 0.05 & 0.22 &  2.471 & 0.224 & 0.000 & 0.000 & 0.534 \\
$1^{+}$ & 3.3(2) & 3.2(9) & 2.8 & 0.4 &  4.740 & .937 & -1.179 & 0.029 & 0.664 \\
($1^{-})$ & --- & --- & --- & --- & 5.548 & 1.664 & 0.000 & 0.000 & 2.827 \\
($2^{-}$) & --- & --- & --- & --- &  12.059 & 0.000 & 3.15 & 0.000 & 0.000 \\
\end{tabular}
\end{ruledtabular}
\end{table*}

\section{New states in light of previous experimental data}
Here we study/assess if the low lying 0$^+_1$ and 2$^+_2$ states are consistent with the available experimental data on $^{8}$B and $^{8}$Li nuclei. The structure of $^{8}$B has been extensively studied in p+$^{7}$Be resonance elastic scattering experiments \cite{Goldberg98,Rogachev01,Angulo03,Yamaguchi09}. In Ref. \cite{Angulo03} the $^{7}$Be+p excitation function of elastic scattering was measured in the c.m. energy range from 0.3 to 0.75 MeV. The new states are at 1.9, 2.5, and 3.3 MeV excitation energies and their influence on the low energy part of the excitation function is very small. In general, the fit to the elastic scattering data does not require the low lying 0$^+_1$ and 2$^+_2$ states. The experimental data in Ref. \cite{Goldberg98,Rogachev01,Yamaguchi09} were fitted with only 1$^{+}$ and 3$^{+}$ states at 0.77 and 2.32 MeV and a 2$^{-}$ state at $\sim3$ MeV. (Presence of a 1$^{+}$ state at $\sim3$ MeV was suggested in Ref. \cite{Goldberg98}.) However, the new states have little influence on the excitation function for elastic scattering. Therefore, the absence of these states in the R-matrix analysis of the elastic scattering data cannot be used as an argument against these states. It is interesting to note that in all three previous measurements \cite{Goldberg98,Rogachev01,Yamaguchi09} the cross section at the resonance energy of the 3$^{+}$ state was measured to be $\approx190$ mb/sr at 180$^{\circ}$. The R-matrix fit to our elastic scattering data produces a lower cross section at 180$^{\circ}$, $\approx$160 mb/sr. This is not surprising because the experimental technique used in previous measurements did not separate elastic from inelastic scattering. Protons from inelastic scattering were contributing to the ``elastic'' excitation functions which resulted in higher cross section values of the measured ``elastic'' excitation functions. 

More experimental information is available regarding the structure of the mirror nucleus, $^{8}$Li. One and two neutron transfer reactions, $^{7}$Li(d,p) \cite{Schiffer1967} and $^{6}$Li(t,p) \cite{Ajzenberg1978} were used to populate states in $^{8}$Li. It is very unlikely that bound states in $^8$Li could have been missed in these experiments. Therefore, the 0$^+_1$ and 2$^+_2$ states are probably above the neutron decay threshold (2.03 MeV) in $^8$Li.

\begin{figure}
\includegraphics[width=1\columnwidth]{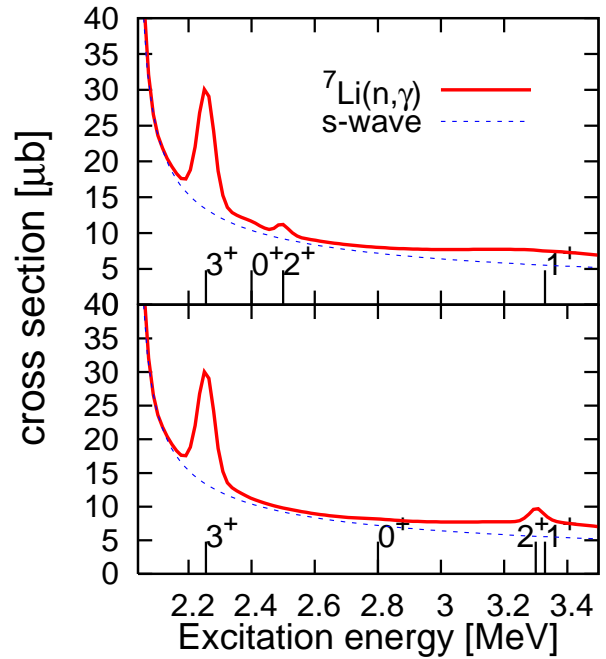} 
\caption{\label{fig:ngamma} (Color online) The $^7$Li(n,$\gamma$) reaction excitation function calculated using TDCSM approach. The known 3$^+$ and 1$^+$ states and the new 0$^+$ and 2$^+$ states at 2.4 and 2.5 MeV  (top panel) and at 2.8 and 3.3 MeV (bottom panel) are shown.
}
\end{figure}

The excitation function for the $^{7}$Li(n,$\gamma$)$^{8}$Li reaction was measured at low c.m. energies (up to 1 MeV) \cite{Imhof1959,Allen1971,Wiescher1989,Blackmon1996}. Only the 3$^+$ state at 2.25 MeV (0.22 MeV above the neutron decay threshold) has been observed. In principle, lack of evidence for the 0$^+$ and the 2$^+$ states in the $^{7}$Li(n,$\gamma$)$^{8}$Li excitation function cannot be considered as a decisive argument against their presence. If the partial $\gamma$ width ($\Gamma_\gamma$) for these states is small then they can be hard to identify within the background from direct neutron capture and resonance capture due to the 3$^+$ state. Fig. \ref{fig:ngamma} shows TDCSM calculations of the (n,$\gamma$) excitation function with the known 3$^+$ and 1$^+$ states and the new 0$^+$ and 2$^+$ states at 2.4 and 2.5 MeV  (top panel) and at 2.8 and 3.3 MeV (bottom panel). It is clear from this figure that observation of the new states in the $^7$Li(n,$\gamma$) reaction is difficult.

\begin{figure}
\includegraphics[width=1\columnwidth]{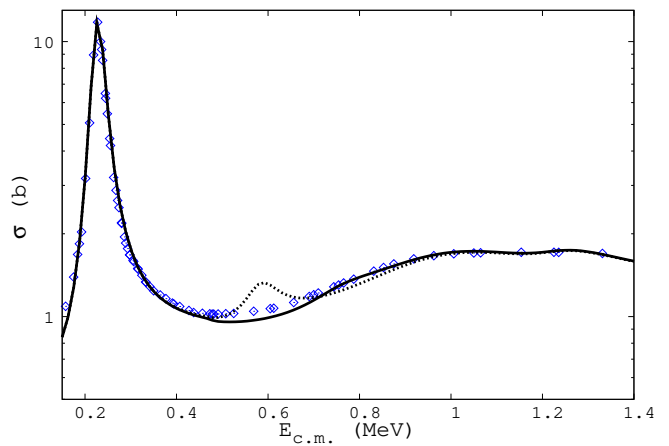} 
\caption{\label{fig:7Linn} Total $^7$Li(n,n)$^7$Li reaction excitation function from \cite{Knox1987}. The solid line is the R-matrix fit with the known 3$^+$ state at 2.25 MeV and the new 0$^+$, 1$^+$ and 2$^+$ states at 2.8, 3.1, and 3.3 MeV.  The dashed line shows the effect of shifting the 0$^+$ state down by 200 keV.}
\end{figure}

Resonances in $^8$Li at excitation energies of up to 9.0 MeV have been studied in elastic and inelastic n+$^{7}$Li scattering and analyzed using the R-matrix approach in Ref. \cite{Knox1987}, where the new low-lying states were suggested. For example, the 0$^+$ state at 3.02 MeV was introduced. Unfortunately, the n+$^7$Li excitation function is relatively featureless, which makes R-matrix analysis ambiguous. The contemporary (for 1987) shell model predictions were used in Ref. \cite{Knox1987} as guidance for the fit. We performed our own R-matrix analysis of the n+$^7$Li excitation functions and attempted to incorporate the new low-lying 0$^+$ and 2$^+$ states into the n+$^7$Li fit. It appears that the low energy n+$^7$Li excitation function for elastic scattering can be reproduced with the 0$^+$, 1$^+$ and 2$^+$ states if they are placed at excitation energies above 2.8 MeV without any modifications to their reduced widths.  The total cross section for the $^7$Li(n,n)$^7$Li(g.s.) reaction is shown in Fig. \ref{fig:7Linn}. The solid line is the R-matrix fit with the states mentioned above, the background states from Table \ref{tab:states}, and the known 3$^+$ at 0.22 MeV.  Note that if the 0$^+$ is shifted down by as little as 200 keV it would appear as relatively narrow peak, which is not observed experimentally (dotted line in Fig. \ref{fig:7Linn}). From the considerations above we can conclude that existence of the new low lying 0$^+$ and 2$^+$ states in $^8$Li does not, in principle, contradict available n+$^{7}$Li elastic scattering experimental data. However, these states have to be shifted up in excitation energy by $\sim$800 keV compared to their suggested location in $^8$B. There is also strong evidence against degeneracy of the new state(s) with the 3$^+$ state. If such degeneracy exists then the experimental cross section at the maximum of the 3$^+$ peak (0.22 MeV) would be higher than can be accounted for by the 3$^+$ state alone. Our R-matrix fit shows that this is not the case.

\begin{figure}
\includegraphics[width=1\columnwidth]{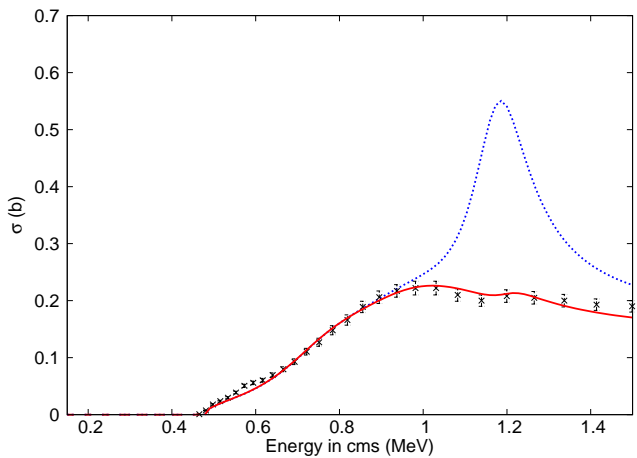} 
\caption{\label{fig:Liinelas} (Color online)  Total $^7$Li(n,n')$^7$Li(1/2-) reaction excitation function from \cite{Olsen1980}.  The dashed blue curve is the R-matrix fit with the reduced width parameters of Table \ref{tab:states}.  The 2$^{+}_{2}$ from the table overestimates the cross section at around 1.2 MeV in the c.m.s.  The solid red curve is the same R-matrix fit, with the elastic component of the 2$^{+}_{2}$ reduced to better fit the data.}
\end{figure}

It is more difficult to reconcile the new states in $^{8}$B and the available $^{7}$Li(n,n')$^{7}$Li(1/2-) experimental data. If reduced widths parameters from Table \ref{tab:states} for these states are used then the $^{7}$Li(n,n')$^{7}$Li(1/2-) cross section is overestimated due to too strong contribution from the 2$^{+}$ state (dotted blue curve in Fig. \ref{fig:Liinelas}). The elastic reduced width amplitude of the 2$^{+}$ state has to be reduced from 0.276 to $<$0.1 in order to produce a good fit to the $^{7}$Li(n,n')$^{7}$Li(1/2-) data (red solid curve in Fig. \ref{fig:Liinelas}). All other parameters for the 2$^{+}$ and also all parameters for the 0$^{+}$ and 1$^{+}$ do not require any modification. The $^{7}$Be(g.s.)+p spectroscopic factor for the 2$^{+}$ state is already small (4$\%$) in $^{8}$B but it appears that it needs further reduction to less than 1$\%$ in $^{8}$Li to reproduce the $^{7}$Li(n,n')$^{7}$Li(1/2$^-$) data.  We do not have a good explanation for this situation.

The excitation energy shift of 800 keV between states in mirror nuclei (Thomas-Ehrman shift \cite{Ehrman1951,Thomas1952}) is very large. While not unique (for example, the shift between the 1/2$^+$ second excited state in $^{19}$O and $^{19}$Na is 730 keV \cite{Angulo2003,Skorodumov2006}) it is generally associated with single particle structure, where the valence nucleon is in the s-wave state. A large Thomas-Ehrman shift results from different asymptotic behavior of the valence nucleon wave function between bound and unbound states in mirror nuclei (Nolen-Schiffer effect \cite{Nolen1969}). The 0$^+$ and 2$^+$ are p-shell states, therefore, a large Thomas-Ehrman shift is not expected. Realizing that this is an unusual situation, we attempted to reproduce the observed p+$^7$Be inelastic scattering excitation function without introducing the new resonances in $^8$B but assuming a direct excitation mechanism of the $^7$Be first excited state in p+$^7$Be scattering. 

\begin{figure}
\includegraphics[width=0.8\columnwidth]{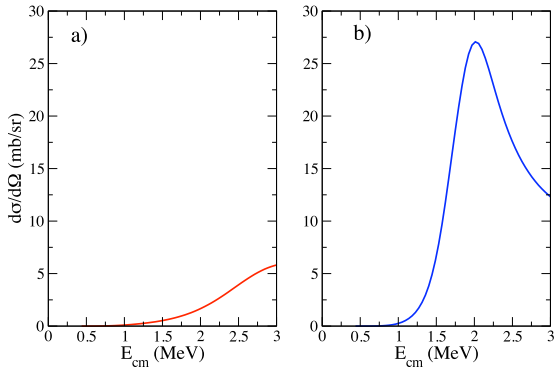}
\caption{ \label{fig:cc} Inelastic scattering differential cross section from the $^{7}$Be(p,p') reaction calculated within the coupled-channels 
approach, assuming a direct mechanism. The left panel uses a p+$^7$Be potential which does not contain resonances within this energy interval. The right panel shows the result of the calculation using a potential that contains a resonance.}
\end{figure}

These calculations were performed using the  coupled-channels approach. The potential of Kim \textit{et al.} \cite{kim87} was first considered for the bare part of the  $^7$Be interaction. The transition potential for the coupling between the ground and the first excited state was generated deforming the bare potential and assuming that these two states of the $^7$Be nucleus are members of a $K=1/2$ rotational band with a quadrupole deformation length of $\delta_2 = 2$~fm. Besides the transition potential, this procedure gives rise also to reorientation terms, which were also taken into account in the calculations. 
The coupled equations were solved to all orders using the computer code {\sc fresco} \cite{fresco}. In Fig.~\ref{fig:cc}a) we show the excitation function for  a $\theta_\mathrm{c.m.}=146^\circ$ as a function of the p+$^7$Be c.m.\ energy. Clearly, the contribution of the direct mechanism is very small 
in this energy range, suggesting that the magnitude of the measured inelastic cross section at these energies cannot be explained by a pure direct reaction mechanism. We performed a second coupled-channels calculation using a potential that produces a resonance at these energies. This potential was parametrized using a Woods-Saxon shape, with radius $R=2.23$~fm (deduced from the matter radius of the $^7$Li nucleus), diffuseness $a=0.65$~fm and the depth adjusted to produce a resonance around $E_\mathrm{c.m.}=2$~MeV. The calculated inelastic excitation function obtained with this potential is given by the solid line in Fig.~\ref{fig:cc}b). The presence of the resonance produces a pronounced maximum about 2~MeV and a significant increase of the magnitude of the cross section. So, based on the coupled-channels analysis we conclude that the high inelastic scattering cross section cannot be reproduced unless resonance(s) is(are) introduced in the corresponding energy range. 

Finally, we have to make an important distinction between the 0$^+$ and the 2$^+$ states. While existence of the 2$^+$ state is hard to dismiss, the case for the 0$^+$ state is somewhat weaker. In spite of the fact that without this resonance the inelastic cross section at 2.0 MeV due to direct excitation of the first excited state in $^7$Be is 3 times smaller than the experimental value, one should be careful making the final call based on such evidence. Further investigation is warranted. Specifically, accurate measurement of the p+$^7$Be excitation function of inelastic scattering in the energy range from 0.7 to 2.0 MeV and in a broad angular range should provide a definitive answer on the existence of the 0$^+$. At this point we can only regard this state as tentative. 

\section{\label{sec:csm} The continuum shell model analysis of the p+$^7$Be data. }
The time dependent continuum shell model \cite{Volya2009} was used as an alternative and more microscopically constrained way to analyze the p+$^{7}$Be data. This model extends the traditional shell model into the domain of reaction physics. It incorporates the many-body dynamics with all essential structure and reaction components, and allows one to predict the reaction observables. Some features, such as the angular dependence of cross sections and interference between resonances are particularly sensitive to the many-body structure. The TDCSM is built upon one of the well-established Hamiltonians of the traditional shell model coupled to reaction channels, where a Woods-Saxon shaped potential is taken from a global Woods-Saxon parametrization \cite{Schwierz2007}. This theoretical treatment of $^{8}$B using the WBP shell model Hamiltonian \cite{Brown2001} is reported in Ref. \cite{Volya2009}. The WBP Hamiltonian was selected because unlike most interactions it results in low-lying $1_{2}^{+},\,0_{1}^{+},$ and $2_{2}^{+}$ states in $^{8}$B, at excitation energies below 3 MeV. To consider a full spectrum of possible Hamiltonians in this investigation, in addition to WBP, we use PWT \cite{Brown2001} and CKI \cite{Cohen1965} shell model interactions. The comparison of the experimental spectroscopic factors of the positive parity states in $^8$B to the predictions of the shell model with different interactions is given in Table \ref{tab:spectroscopic}. The experimental spectroscopic factors were calculated as the ratio between the partial width and the single particle width calculated using a Woods-Saxon potential with a global Woods-Saxon parametrization \cite{Schwierz2007}.  In Table \ref{tab:spectroscopic} it can be seen that all three residual interactions are in good agreement with the experimental spectroscopic factors for the 1$^{+}_{1}$ and 3$^{+}_{1}$ resonances.  All three interactions reproduce the inelastic spectroscopic factor for the 0$^{+}_{1}$, and both the elastic and inelastic components of the 1$^{+}_{2}$. The WBP interaction is the only interaction that predicts a 2$^{+}_{2}$ that is dominated by an inelastic component, as is seen experimentally, while the PWT and CKI interactions both predict a 2$^{+}$ state with a similar inelastically dominated component as the 2$^{+}_{3}$.  

\begin{figure}
\includegraphics[width=0.8\columnwidth]{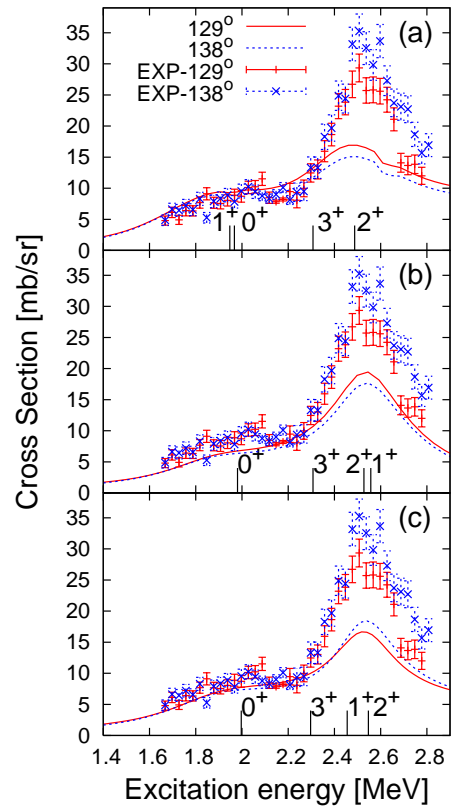}
\caption{ \label{fig:TDCSM} (Color online) Inelastic scattering differential cross sections for $^{7}$Be(p,p') reaction obtained with the TDCSM that uses three different Hamiltonians is compared to the experimental data. The panels (a), (b), and (c) correspond to WBP, PWT, and CKI interactions.}
\end{figure}

\begin{table}
\caption{\label{tab:spectroscopic} Experimental spectroscopic factors of positive parity states compared to the shell model predictions.}
\begin{ruledtabular}
 \begin{tabular}{ccccccccccccccccccccc}
 J$^\pi$                &   1$^+$ & 0$^+$ & 3$^+$ & 2$^+$ & 1$^+$ \\
 E$_{ex}$ (MeV)         & 0.768  & 1.9      & 2.31   & 2.50   & 3.3    \\
 \hline
S$_{^7Be(g.s.)+p}$$^{*}$  & 0.38 & 0.05 & 0.20 & 0.04 & $\approx$1 \\
S$_{^7Be(1/2^-)+p}$$^{*}$ & --- & 0.94 & --- & 0.19 & 0.14 \\
\hline
		    &1$^{+}_{1}$&0$^{+}_{1}$&3$^{+}_{1}$&2$^{+}_{2}$&1$^{+}_{2}$&2$^{+}_{3}$\\
E$_{CKI}$           & 1.08 & 4.95 & 1.69 & 4.24 & 2.77 & 5.15 \\
CKI el.  & 0.44 & 0.34 & 0.33   & 0.56 & 1.10  & 0.06\\
CKI in.  & 0.87 & 0.90 &   ---     & 0.04 & 0.14 &  0.27\\
\hline
E$_{PWT}$  & 1.54 & 4.01 & 2.14 & 4.39    & 3.80  & 6.06\\
PWT el.  & 0.45   & 0.27 & 0.30   & 0.55   & 0.95   & 0.11   \\
PWT in.  & 0.84   & 0.96 &   ---      & 0.03   &   0       & 0.22   \\
\hline
E$_{WBP}$  & 0.55 & 1.75   & 1.99   & 2.40  & 1.73 & 3.25\\
WBP el.  & 0.40 & 0.48 & 0.37   & 0.13   & 1.0     & 0.40 \\
WBP in.  & 0.77 & 0.84 &   ---      & 0.40   & 0.14   & 0.03 \\
\end{tabular}
\end{ruledtabular}

$^{*}$ Experimental values.
\end{table}

The best validation of the theoretical model predictions can be performed if the measured cross section is calculated directly from the model. Unfortunately, the reaction physics is very sensitive to kinematics and to the exact position of levels in the spectrum because of the phase space and barrier penetrability. While the traditional shell model may, in general, be good in describing positions and ordering of states, often its precision is not close to what is required by the reaction physics. Thus, it is common practice to set the exact reaction kinematics based on observation. In our approach all known states and thresholds are adjusted from experimental data and we treat the energies of unknown $1_{2}^{+},\,0_{1}^{+},$ and $2_{2}^{+}$ states as parameters. In our study we vary these three parameters to best fit the observed cross section. The TDCSM provides an effective mechanism to modify the position of any state in the Hamiltonian while keeping all structural aspects unchanged. This is done by adding to a shell model Hamiltonian a factorisable term $\delta E|\alpha\rangle\langle\alpha|$, where $|\alpha\rangle$ is the eigenstate to be shifted and $\delta E$ is the energy shift. The corresponding change in the many-body propagator is performed exactly with the help of Dyson's equation, for details see Ref. \cite{Volya2009}. 

In Fig. \ref{fig:TDCSM} the inelastic scattering cross section for $^{7}$Be(p,p') obtained with TDCSM is compared to experiment. Panels (a), (b), and (c) correspond to calculations with WBP, PWT, and CKI interactions, respectively. The spin and parities of resonances in the energy region plotted are marked. The inelastic cross section is not sensitive to the $3^{+}$ state which is seen in the elastic scattering cross section. All models predict a similar structure of the $3^{+}$ state and therefore produce a comparable elastic cross section which agrees well with experiment. The elastic $^{7}$Be(p,p) cross section with WBP interaction is demonstrated in Ref. \cite{Volya2009}. Positions of $1_{2}^{+},\,0_{1}^{+},$ and $2_{2}^{+}$ resonances, indicated in Fig. \ref{fig:TDCSM}, are not known {\it a priori}; here they are adjusted by visual examination to best reproduce the experimental data. The main peak in the $^{7}$Be(p,p') cross section is due to the $2_{2}^{+}$ resonance at around 2.5 MeV of excitation. It was found that agreement with the experimental data is good if the $0_{1}^{+}$ is placed around $2$ MeV and the $1_{2}^{+}$ is moved to higher excitation energy.
(Sensitivity of the inelastic cross section to the position of the $1_{2}^+$ state is weak. However, the $1_{2}^{+}$ state at excitation energies below 2.3 MeV would produce a peak in the elastic cross section, which is not observed experimentally. See discussion in Section \ref{sec:Rmatrix}.) In the case of the WBP interaction, Fig. \ref{fig:TDCSM}(a), no position adjustment was made to the $1_{2}^{+}$ and $0_{1}^{+}$ states and the $2_{2}^{+}$ is only moved down by about 140 keV. The CKI Hamiltonian gives two $2^{+}$ excited states at 4.2 and 5.1 MeV of excitation. Both of these states have been tried as candidates for the 2.5 MeV resonance and it was determined that the second 5.1 MeV state in the CKI Hamiltonian has the correct structure. Our main conclusion from the calculations shown in Fig. \ref{fig:TDCSM} is that the CKI interaction appears to be best in reproducing the cross section. The states obtained with the CKI appear to have structure which agrees with the observed interference and angular dependence features. In particular, only the CKI interaction is able to reproduce the observed increase in the cross section at 2.5 MeV for higher angle (Fig. \ref{fig:TDCSM}(c)). The height of the resonance peak at 2.5 MeV is the primary difference between theory and observation. We attribute this difference partially to the 1$^-$ state, which was not included in the shell model analysis (only p-shell states were considered) and also to the somewhat different ratio between the elastic and inelastic partial widths for the 2$^{+}_{2}$.

The amplitudes from the final R-matrix fit with the CKI interaction are summarized in Tab. \ref{tab:Decay-amplitudes} (Resonance reduced width parameters from the R-matrix fit were re-coupled from the LS to the jj coupling scheme for direct comparison with TDCSM amplitudes). It should be noted that the choice of channel radius in the R-matrix calculations will have the affect of scaling the reduced width parameters, thus one should not directly compare absolute values, but rather sign and relative values of the reduced widths.  We note that the fit only slightly modifies the amplitudes for the $3_{1}^{+}$ and $2_{2}^{+}$ states leaving the general features of CKI unchanged. However, there is a significant difference between experimental excitation energies and CKI predictions for the $2_{2}^{+}$ state.  The $0_{1}^{+}$ also has a significant shift between the experimental and CKI predicted energy.


\begin{table*}
\caption{\label{tab:Decay-amplitudes} Decay amplitudes from the final R-matrix fit and from the CKI based TDCSM. The first column denotes the spin and parity of the resonance, following are excitation energy and four amplitudes from the R-matrix fit. Excitation energy and amplitudes for states in $^{8}$B from the CKI Hamiltonian are listed in the remaining five columns. }
\begin{ruledtabular}
\begin{tabular}{cc|cc|cc|c|cc|cc}
\multicolumn{6}{c|}{R-matrix fit} & \multicolumn{5}{c}{TDCSM with CKI interaction}\tabularnewline
\hline
 &  & \multicolumn{2}{c|}{$^{7}$Be 3/2$^{-}$ g.s.} & \multicolumn{2}{c|}{$^{7}$Be 1/2$^{-}$} &  & \multicolumn{2}{c|}{$^{7}$Be 3/2$^{-}$ g.s.} & \multicolumn{2}{c}{$^{7}$Be 1/2$^{-}$}\tabularnewline
 \hline
$J^{\pi}$ & E{[}MeV] & p$_{1/2}$ & p$_{3/2}$ & p$_{1/2}$ & p$_{3/2}$ & E{[}MeV] & p$_{1/2}$ & p$_{3/2}$ & p$_{1/2}$ & p$_{3/2}$\tabularnewline
\hline
2$_{1}{}^{+}$ & 0    & 0.19    & -0.94 &         & 0.43  & 0.00 & 0.23  & -0.98 &       & -0.43\tabularnewline
1$_{1}{}^{+}$ & 0.768 & -0.17  &  0.71 &  0.24   & -0.91 & 1.08 & -0.35 & 0.57  & 0.24 & -0.91\tabularnewline
0$_{1}{}^{+}$ & 1.9  &          &  0.35 & 1.30   &        & 4.95 &        & -0.59  & 0.95 &      \tabularnewline
3$_{1}{}^{+}$ & 2.31 &          &  0.61 &         &        & 1.69 &        & 0.58 &       &      \tabularnewline
2$_{2}{}^{+}$ & 2.50 & -0.17  & 0.17  &         & 0.53  & 5.15 & -0.17 & 0.16  &       & -0.52 \tabularnewline
1$_{2}{}^{+}$ & 3.3  & -1.46   & 0.37  & 0.53  & 0.41  & 2.77 & 0.84 & 0.62 & 0.33 & 0.18 \tabularnewline
\hline
\end{tabular}
\end{ruledtabular}
\end{table*}

\section{Comparison of the {\it ab initio} models to the new experimental data. \label{abinitio}}
The first {\it ab initio} calculations for A=8 isotopes were performed in 1998 by P. Navratil and B. Barrett \cite{Navratil98} using the large-basis no-core shell model approach. Except for the broad 1$^+$ state at 3.2 MeV (which was under-bound by 2 MeV) the known states have been well reproduced by the calculations. In addition to the known states the new 0$^+$ and 2$^+$ states have been predicted at 4.5 and 5.0 MeV respectively. While these states had rather high excitation energy they were still the third and the fourth excited states in the calculations, below the second 1$^+$ state (predicted at 5.2 MeV). Refined NCSM calculations with CD-Bonn 2000 potential \cite{Navratil2006} produced the 0$^+$ as the second excited state at 2.23 MeV in $^8$B and 2.48 MeV in $^8$Li. The 2$^+$ state was produced at 3.8 MeV. This prediction is remarkably close to the experimental result of this work with respect to the 0$^+$ state, suggested at 1.9 MeV in $^8$B. Another interesting prediction is that the excitation energy of this state should be shifted up in $^8$Li by 250 keV. Much smaller shifts were predicted for the known 1$^+$ and 3$^+$ states (50 and 40 keV respectively).

Similarly to the NCSM, the new low-lying positive parity states were predicted by the Green Function Monte Carlo method \cite{Wiringa00}. The 0$^+$ state was predicted as the second excited state with excitation energy of 1.91(29) MeV (the results were given only for $^8$Li). The GFMC results for the 2$^+$ state are not available, nevertheless, the variational Monte Carlo calculations produce this state at 3.86(18) MeV \cite{Wiringa00}. However, the more recent version of GFMC calculations \cite{Pieper04} with the AV18/IL2 Hamiltonian puts the 0$^+$ and 2$^+$ at excitation energies of 3.6 and 5.3 MeV, respectively. (This version of the GFMC calculation also puts the known 1$^+$ state at 4.7 MeV instead of 3.2 MeV.) 

Clearly, there is no unified picture for the level structure of the $^8$B - $^8$Li isotopes from the available array of {\it ab initio} calculations. However, all of them produce a 0$^+$ as either the second or third excited state, always below the known 1$^+_{2}$ state (experimentally found at 3.2 MeV in $^8$Li). Our experimental result confirms this prediction. The 2$^+$ state is generally found at higher excitation energy in {\it ab initio} calculations than observed in this work for $^8$B.

\begin{figure}
\includegraphics[width=0.8\columnwidth]{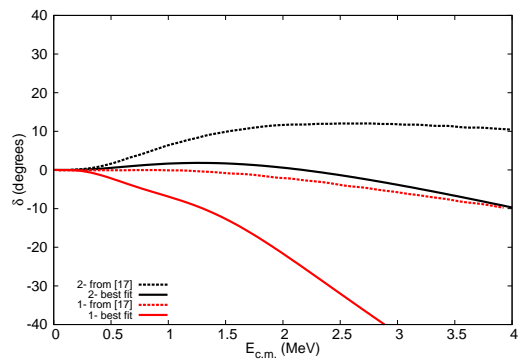}
\caption{ \label{fig:negphase} (Color online) Phase shifts from the R-matrix best fit and from Navr\'{a}til \cite{Navratil10} for the negative parity states. R-matrix calculations are the solid black and red curves for the 2$^{-}$ and 1$^{-}$ respectively.  The calculated {\it ab initio} phase shifts from \cite{Navratil10} are the dashed black and red curves for the 2$^{-}$ and 1$^{-}$ respectively. Note that the sensitivity of the fit to the 1$^-$ phase shift is very weak as discussed in Chapter \ref{abinitio}.}
\end{figure}

\begin{figure*}
\includegraphics[width=0.8\textwidth]{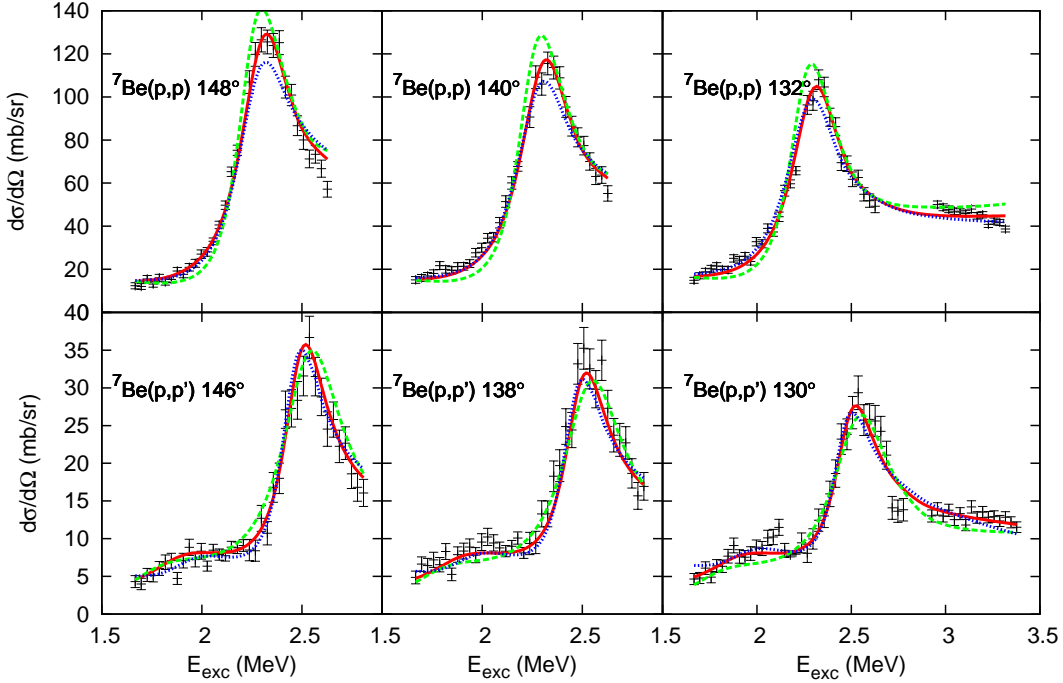}
\caption{ \label{fig:totalours} (Color online) The excitation functions for the $^7$Be+p elastic and inelastic scattering. The best fit is the solid red curve.  The dashed green line is the R-matrix fit with a 2$^{-}$ phase shift matching the work of \cite{Navratil10}. The blue dotted curve is the fit with the 1$^{+}$ phase shifts matching the phase shifts from \cite{Navratil10}.}
\end{figure*}

With the development of the {\it ab initio} NCSM/RGM approach \cite{Navratil11}, the {\it ab initio} phase shifts can now be compared directly to the experimental phase shifts extracted from the R-matrix analysis of the experimental data. The $^7$Be+p diagonal phase shifts as well as $^7$Be(p,p') excitation function have been calculated in Ref. \cite{Navratil10}. Fig. \ref{fig:negphase} shows the {\it ab initio} negative parity 1$^{-}$ and 2$^{-}$ phase shifts from \cite{Navratil11} as red and black dashed curves respectively. The experimental 1$^{-}$ and 2$^{-}$ phase shifts from the R-matrix best fit are the red and black solid curves respectively. The experimental and the theoretical phase shifts appear to be different. However, the general trend is reproduced. Moreover, the fit is not very sensitive to the 1$^-$ phase shift. It is possible to make a good fit with $\chi^2$=0.92 using the 1$^-$ phase shift from \cite{Navratil10}. While it is marginally worse than the best fit $\chi^2$ (0.89) it does not differ significantly visually and the parameters for all other states remain within uncertainty quoted in Table \ref{tab:states}. Therefore, the sensitivity of the fit to the 1$^-$ phase shift is small and seemingly large difference shown in Fig. \ref{fig:negphase} should not be considered as significant. The fit is more sensitive to the 2$^-$ phase shift. It does not reach the maximum value of 12 degrees predicted by the NCSM/RGM. If the 2$^-$ phase shift from \cite{Navratil11} is used in the R-matrix fit then the $\chi^2$ of the best fit is 2.2 and the fit is visually worse. This is shown in Fig. \ref{fig:totalours} by the dashed green curve.

\begin{figure}
\includegraphics[width=1.0\columnwidth]{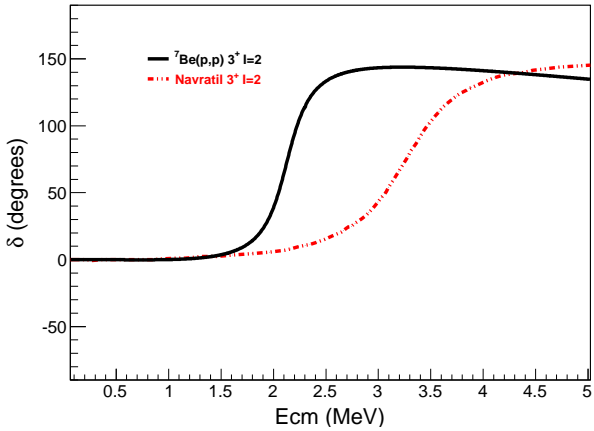}
\caption{ \label{fig:3+phase} (Color online) Phase shifts from our R-matrix best fit for the 3$^{+}$ (solid black curve) and calculations of Navr\'{a}til (dashed dotted red line).}
\end{figure}

The shape of the 3$^+$ diagonal phase shift is determined by the 3$^+$ state at 2.29 MeV, which only has contribution from channel spin S=2 and no inelastic component. NCSM/RGM overestimates the excitation energy of this state by $\sim$1 MeV as can be seen in Fig. \ref{fig:3+phase}, otherwise the phase shift would be similar to the experimental one. This was verified by shifting the experimental excitation energy of the 3$^+$ state to the NCSM/RGM result in the R-matrix calculations.

\begin{figure}
\includegraphics[width=1.0\columnwidth]{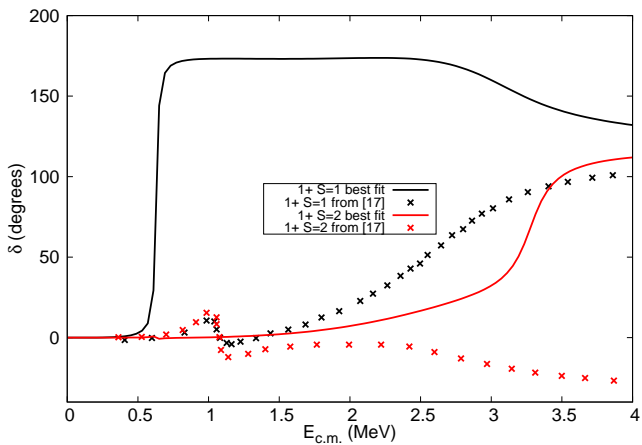}
\caption{ \label{fig:1+phases} (Color online) The 1$^{+}$ phase shifts from the best fit and from \cite{Navratil10}.  Best fit R-Matrix phase shifts for channel spins 1 and 2 are black and red solid curves respectively. The black and red crosses are S=1 and S=2 1$^+$ phase shifts from \cite{Navratil10}.}
\end{figure}

Comparison of the 1$^+$ phase shifts from the best fit and from the \cite{Navratil10} is shown in Fig. \ref{fig:1+phases}. The phase shifts appear to be very dissimilar. The 1$^+_1$ state shows up predominantly in the S=1 channel in the best fit (and also in the shell model calculations and in \cite{Angulo03}) and it is located at lower energy than predicted in \cite{Navratil10}. This makes the contribution from the inelastic channel negligible and the S=1 phase shift goes through 90 degrees, unlike in \cite{Navratil10}. More important difference is that the best fit S=2 phase shift barely shows any sign of the 1$^+_1$ state while the {\it ab initio} S=2 and S=1 phase shifts have about equal contribution from the 1$^+_1$ state. In principle, a good fit can be achieved with the phase shifts similar to those calculated in Ref. \cite{Navratil10,Navratil11} (provided that the first excited state is shifted down by about 300 keV from where it appears in the {\it ab initio} calculations). All excitation functions are reproduced with this solution, except for the high and low angles in low-energy data from \cite{Angulo03}. This is shown in Fig. \ref{fig:angulofig}. We believe this discrepancy is significant and favors the CKI shell model prediction over the {\it ab initio} calculations for the structure of the 1$^+_1$ state in $^8$B.

The 1$^+_2$ state is responsible for the behavior of the 1$^+$ phase shifts  above 1 MeV. Comparing the best fit and {\it ab initio} phase shifts one can notice that at higher energies the best fit S=1 phase shift is similar to the S=2 {\it ab initio} phase shift and the best fit S=2 phase shift is similar to the S=1 {\it ab initio} phase shift. We have produced another fit using the {\it ab initio} 1$^+$ phase shifts from \cite{Navratil10} and varied parameters for all other states. This fit has the $\chi^2$ at 1.2. The fit to the inelastic data is visually identical. Quality of the fit to the elastic scattering data is somewhat worse (see Fig. \ref{fig:totalours}). Observable parameters for all other positive parity states were still within the uncertainties quoted in Table \ref{tab:states}. Generally, while the best fit 1$^+$ phase shifts look very different, we find that the data can be reproduced reasonably well with the 1$^+$ phase shifts from  \cite{Navratil10}. This ambiguity can be resolved if wide range of angles and energy is measured with high accuracy and/or experiment is performed with the polarized target. 

\begin{figure}
\includegraphics[width=1.0\columnwidth]{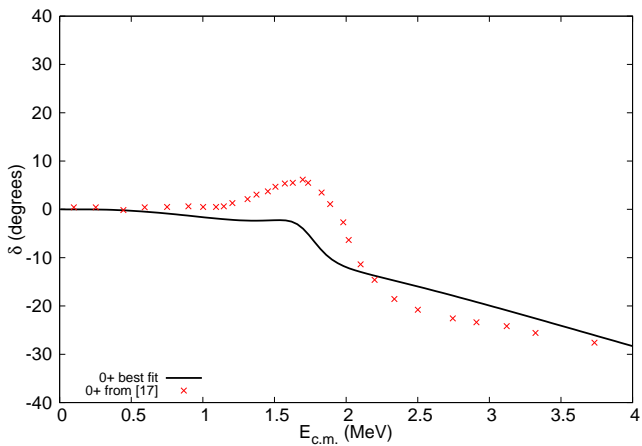}
\caption{ \label{fig:0+match} (Color online) The 0$^{+}$ phase shifts from the best fit (solid black curve) and from \cite{Navratil10} (red crosses).}
\end{figure}

The 0$^+$ phase shift is defined by the $0^+$ resonance at 1.9 MeV. Fig. \ref{fig:0+match} shows the 0$^+$ phase shifts from the best fit and the {\it ab initio} calculations \cite{Navratil10}. The two phase shifts are very similar, indicating that the structure of the 0$^+$ state is well reproduced in \cite{Navratil10}. In order to have a perfect match between our phase shift and that from \cite{Navratil10} it is necessary to increase the total width of the state to $~\sim$1 MeV. A fit to the experimental data with the  0$^+$ phase shift from \cite{Navratil10} produces a $\chi^2$ value of 0.97 and is almost indistinguishable visually. This is because the stronger 0$^+$ is compensated by the modifications to the negative parity shifts and the parameters for the positive parity states remain almost unchanged. This is why the 0$^+$ state has large uncertainly for its widths (see Table \ref{tab:states}).

\begin{figure}
\includegraphics[width=1.0\columnwidth]{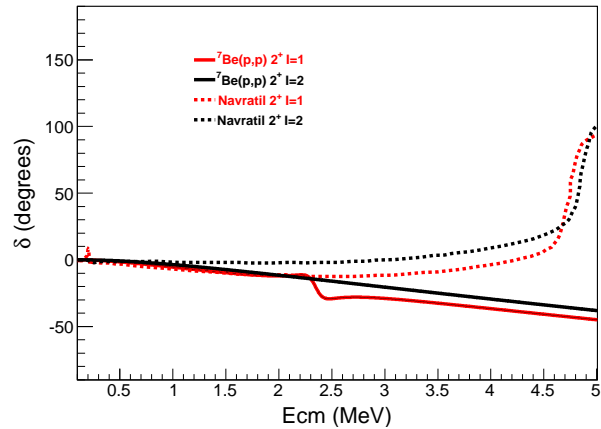}
\caption{ \label{fig:2+phases} (Color online)  Comparison of the experimental best fit 2$^{+}$ phase shifts to the calculated phase shifts of \cite{Navratil10}.  R-Matrix calculations for channel spins 1 and 2 are red dashed-dotted and black dashed-dotted curves respectively, while those of \cite{Navratil10} are solid red and black curves respectively.}
\end{figure}

The only obvious difference between the experimental data and the results of NCSM/RGM calculations \cite{Navratil10,Navratil11} is related to the 2$^+$ phase shift. The 2$^+_2$ state is predicted at 5 MeV by the NCSM/RGM calculations and has about equal contribution from both channel spins. The R-matrix fit to the observed 2$^+$ state favors channel spin S=1. The dominant decay mode for the experimental 2$^+_2$ state is into the first excited state of $^7$Be(1/2$^-$). This produces the characteristic shape of the S=1 2$^+$ phase shift that is very different from the {\it ab initio} phase shifts as seen in Fig. \ref{fig:2+phases}. It appears that the 2$^+$ state predicted in Ref. \cite{Navratil10,Navratil11} and the observed 2$^+$ are two different states. We can speculate that the situation here may be similar to the predictions of the conventional shell model CKI Hamiltonian, that produces two 2$^+$ states at 4.2 and 5.1 MeV and only the latter has the correct structure (see also discussion in Section \ref{sec:csm}). It is possible that the lowest 2$^+$ state predicted by the NCSM/RGM calculations is not the one observed in this experiment.

\section{Conclusion}
The excitation function for p+$^7$Be elastic and inelastic scattering was measured in the energy range of 1.6 - 3.4 MeV and angular range of 132 - 148 degrees. An R-matrix analysis of the excitation functions indicates that new low-lying states in $^8$B have to be introduced in order to explain the large inelastic cross section with a well-defined peak at 2.5 MeV. These new states are suggested to be the 0$^+$ at 1.9 MeV, 2$^+$ at 2.5 MeV, and 1$^{+}$ at 3.3 MeV with width 530 keV, 270 keV, and 3.2 MeV respectively. Evidence for the 2$^+$ state at 2.5 MeV is reliable. The 1$^{+}$ is needed to reproduce the high-energy inelastic cross section and is seen in the mirror nucleus $^{8}$Li. The 0$^+$ at 1.9 MeV can be considered as tentative and more accurate measurements are needed, especially at the lower excitation energy region. However, uncertainty is not related to the spin-parity assignment. If there is a state at 1.9 MeV then it has to be the 0$^+$, as any other spin-parity assignment does not allow fits to the elastic and inelastic scattering simultaneously. The uncertainty is related to the possibility of explaining the observed enhancement of the inelastic scattering cross section at energies below 2.3 MeV by direct excitation of the first excited state in $^7$Be+p scattering. Coupled-channels calculations of the $^7$Be(p,p') inelastic scattering cross section assuming a direct mechanism have been performed and it was shown that without resonance(s) the cross section is significantly smaller than observed experimentally. Nevertheless, taking into account uncertainties of the coupled-channel calculations we consider the 0$^+$ state as tentative.

Analysis of the available experimental data on the mirror nucleus, $^8$Li, indicates that it is unlikely that the new states can be found at excitation energies below 2.8 MeV in $^8$Li. Therefore, the excitation energy shift for these new states between the two mirror nuclei is $\sim$800 keV. This is a factor of two or three larger than the typical Thomas-Ehrman shift in p-shell nuclei. We cannot offer an explanation for this phenomenon.

The Time Dependent Continuum Shell Model provided important guidance in the R-matrix analysis of the experimental data. Due to the presence of many broad overlapping resonances the number of free parameters is large and a blind R-matrix fit is ambiguous. Therefore, theoretical constraints become very important for extracting more reliable results from the R-matrix fit. This is a typical situation for the resonance scattering with exotic nuclei and development of the theoretical tools for this problem is an important step forward.

The recent development of {\it ab initio} NCSM/RGM calculations open up an exciting opportunity for a sensitive tests of {\it ab initio}  models. We performed a detailed comparison of the diagonal p+$^7$Be phase shifts calculated in Ref. \cite{Navratil10,Navratil11} to the experimental data. Overall, with the exception of the 2$^+$ phase shift the results are encouraging and the predictions of the {\it ab initio} model are close to the R-matrix best fit.

The work was supported by National Science Foundation under grant number PHY-456463 and by the U.S. Department of Energy grant DE-FG02-92ER40750.

\bibliography{8B.bib}

\end{document}